\documentclass[useAMS,onecolumn]{raa}
\usepackage{graphicx}
\usepackage{natbib}

\begin{document}

\title {Searching Signals in Chinese Ancient Records for the
$^{14}$C Increases in AD 774-775 and in AD 992-993}

\author{Ya-Ting Chai\inst{1}, and Yuan-Chuan Zou\inst{1} 
 }
\institute{School of Physics, Huazhong University of Science and Technology, Wuhan 430074, China; {\it zouyc@hust.edu.cn(YCZ)}
}
\date{\today}

\abstract{
According to the analysis of the $^{14}$C content of two Japanese trees
over a period of approximately 3000 years at high time resolution, \citet{Miyake2012} found a rapid increase 
at AD 774-775 and later on at AD 992-993 \citep{Miyake2013}. This corresponds to a high-energy event happened within
one year that input $\gamma$-ray energy about 7$\times{}$10$^{24}$erg to the Earth, leaving the origin a mystery.
Such strong event should have an unusual optical counterpart, and have been recorded  in historical
literatures. We searched Chinese historical materials around AD 744-775 and AD 992-993, but no remarkable event was found except a violent thunderstorm in AD 775. However, the possibility of a thunderstorm containing so much energy is still unlikely. We conclude the event caused the $^{14}$C increase is still unclear. This event most probably has no optical counterpart, and short gamma-ray burst, giant flare of a soft gamma-ray repeater and terrestrial $\gamma$-ray flash may all be the candidates.
}

 \authorrunning{Y. T. Chai \& Y. C. Zou}
 \titlerunning{Chinese Ancient Records for the $^{14}$C Increases}
 \maketitle
 
\section{\label{sec:intro}Introduction}
Two Japanese cedar trees recorded the significant increase of 1.2\% in $^{14}$C content
in AD 774-775 \citep{Miyake2012}. This result is also consistent with IntCal198 \citep{Miyake2012}, which is derived
from diverse trees, such as the Irish oak, German oak and pine \citep{Stuiver1998}. The
$^{10}$Be content also increased by 30\%, that can be extracted in the layers of ice or snow in
Antarctica around AD 775. According to the simulation of the
temporal variations of $^{14}$C content, the high energy event happened within
one year. It is an interesting mystery that what kind of phenomenon released so huge amount of energy to the earth and caused the increase of $^{14}$C and $^{10}$Be.

\cite{Hambaryan2012} and most recently \citet{Pavlov2014} proposed that a galactic short gamma-ray burst (GRB) led
to the increase in $^{14}$C content in AD 774-775.
However, the short GRB was supposed to be located at 1 to 4 kpc, and the rate (directed to the Earth) to be roughly once per 1300 years, which is much larger than the estimated rate -- about 1000 Gpc$^{-3}$ yr$^{-1}$ (taking the beaming factor 25, and a typical luminosity $L\sim 10^{49} {\rm erg\, s^{-2}}$) \citep{Nakar2007}.
 On the other hand, there should be no short GRBs located inside 1 kpc in the last more than millions of years, as so nearby GRBs may destroy the life on earth \citep{Melott2004,Thomas2005}. The inconsistency on the rate make the short gamma-ray burst unlikely.
Later on, \citet{Miyake2013} found another $^{14}$C increase event in AD 992-993 at the measurement from AD 822 to 1020. This discovery notably increases the event rate, which may make the GRB model less likely.

\begin{figure}
  \centering
  \includegraphics[angle=0,width=0.5\textwidth]{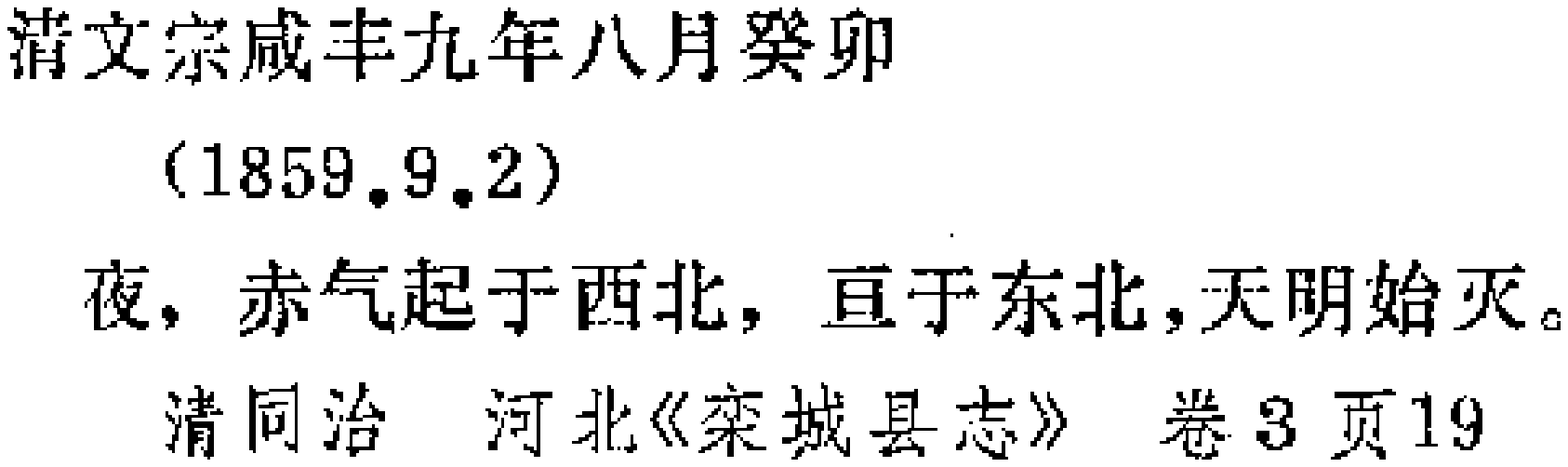}
  \caption{The record of aurora in 1859 from {\it The General Catalogue of Chinese Ancient Astronomical Records} \citep{Bao1988}, which copied this event from a historical book {\it Luancheng Xianzhi}. This aurora was related to the Carrington event.}
  \label{fig:1859}
\end{figure}

Looking back to the Sun,  \cite{Melott2012} and \cite{Thomas2013} suggested a huge solar proton event might be the source, while the released energy should be 10 or 20 times
stronger than the Carrington event \citep{Thomas2007}.
However, \citet{Cliver2014} claimed that AD 775 is in an inactive interval of solar activity, which is hard to produce the required solar particle event.
The Carrington event occurred  on 1 September, 1895, was regarded as the
greatest sun activity in the last 200 years. It brought great
influence to the geomagnetic field. In the night, the colourful polar auroras
in the sky had been permeated south to Cuba and Hawaii. The aurora was also recorded in Chinese history book \citep{Bao1988}, as shown in figure \ref{fig:1859}. Consequently, the solar event at AD 774-775 might also have produced even stronger polar aurora. However, there was no record about that period, which will be shown in section 2.

As pointed by \citet{Miyake2012} that the event was global, the huge amount of $\gamma$-ray energy ($\sim 10^{25}$ ergs to the Earth) released at that time, would most probably have produced some observable phenomena, which might be astronomical event, and could have been recorded. China has a long historical record in the literatures. They are mostly written  by the official organizations, which assure the reliability. We searched the records on the Chinese ancient literatures around AD 774-775 and AD 992-993, to find whether there was counterpart recorded. Though we didn't find any astronomical counterpart in the historical records, the details of the records several years around AD 774-775 are listed in section 2, while leaving AD 992-993 alone; in section 3, we discuss a probable event found from the literatures, the terrestrial $\gamma$-ray flashes (TGFs); and conclude in section 4.

\section{Historical Records}

Chinese historical literatures contain comprehensive data including
historical events, as well as astronomical phenomena. The event leading to the
significant $^{14}$C increase might also have produced other observable phenomena on
the Earth. AD 775 was in the 10th year of the \textit{Dali} reign period at the Tang dynasty of China. We then searched numerous
Chinese historical literatures about that period. These books are:
\textit{Jiutangshu} \citep{Liu945}, finished at the Later Jin Dynasty period, which is an officially edited history book about Tang Dynasty for AD 618-907, where two issues (issue 35, 36) record specially the astronomical phenomena;
\textit{Xintangshu} \citep{Ouyang1060}, finished at Song Dynasty, which is also an official history book for Tang Dynasty, but contains more abundant materials, as well as the astronomical phenomena;
\textit{Zizhitongjian} \citep{Sima1085}, which is a comprehensive history book about the period of 403 BC -- AD 959;
and several modern books and articles like
\textit{The General Catalogue of Chinese Ancient Astronomical Records} \citep{Bao1988}; \textit{A New Catalogue of Ancient Novae} \citep{Xi1955};
and \textit{Chinese Astronomical History} \citep{Chen2006}.

We list all the records during AD 770-780, and some other remarkable events near that period in the following. As for each event, there may be duplicate records in different books. We just list a few representatives.
\textit{ The General Catalogue of Chinese Ancient Astronomical Records} \citep{Bao1988} has the most comprehensive records, which are all collected from numerous ancient books. We priorly chose records from this book.
This book \citep{Bao1988} records more than 270 sunspots, 300 polar auroras, 300 meteorolites, 1600 solar eclipses, 1100 lunar eclipses, 200 phenomena of lunar occultations of the Planets, 100 novae and supernovae, 1000 comets, 400 meteor showers, and 4900 meteors,
from \textit{the Spring and Autumn period} (BC 770-221) to AD 1991. However,  during AD 770-780, except  a few comets and meteors recorded, there were no record of sunspots, auroras, meteorites, notable meteor showers, nova or supernova. We also did not find records from other literatures.

Here we list the few comets recorded. On 26 May, 770 \citep[][on page 408]{Bao1988}, ``A comet from \textit{Wuche}, (\textit{Wuche}, or \textit{Five chariots}, an ancient Chinese asterism, belongs to Auriga and Taurus), as long as five \textit{Zhang} (1 \textit{Zhang}=3.3333 meters), disappeared on 25 July, AD 770. (\textit{Tanghuiyao}, vol. 43, p. 167); A comet from \textit{Wuche}, radiated brilliant light, as long as three \textit{Zhang}. On 15 June, at the north, it was white. On 19 June, gradually moved to the east, near to the middle star of \textit{Bagu} (\textit{Bagu},  or  \textit{Eight kinds of crops}, an ancient Chinese asterism, belong to Auriga), then moved to \textit{Sangong} (\textit{Sangong}, or  \textit{Three excellencies}, an ancient Chinese asterism, belong to Canes Venatici) on 9 July, and disappeared on 25 July (\textit{Xintangshu.Tianwener}, vol.32, p. 838).'' There are also 4 other records in different books with a similar description, which are not listed here.  There was another comet on 17 January, 773 \citep[][on page 408]{Bao1988}, ``Long star from \textit{Shen} (\textit{Jiutangshu.Daizong}, vol. 11, p. 301, \textit{Xintangshu.Daizong}, vol. 6, p. 176); There was a long star under the \textit{Shen}(\textit{Shen} is one of the lunar mansions, also know as Orion in modern astronomy.), extended to the whole sky. It was a long star, which belongs to a comet (\textit{Xintangshu.Tianwener}, vol. 32, p. 838)". Other similar records are not listed here. However, comets are very unlikely to be the origin of the $^{14}$C increase, and the time is not consistent with the $^{14}$C event.

Several meteors records are listed in the following:
On 2 November, AD 771 \citep[][on page 642]{Bao1988}, ``in the night, a meteor in the southwest, as big as a \textit{Shengqi} (a container with volume about 1 litre), with tail. The light illuminated the ground, like pearl scattering, longer than five \textit{Zhang}. Started from \textit{Xunv} (\textit{Xunv}, or  \textit{Girl}, is one of the lunar mansions, including four stars located in Aquarius in modern astronomy.) and disappeared in the south of \textit{Tianshiyuan} (\textit{Tianshiyuan} is one of three enclosures.) (\textit{Jiutangshu.TianwenXia} vol. 36, p.1327)."
On 18 July, 773, ``There was a meteor as big as \textit{Shengqi}, with a tail about 3 \textit{Zhang}. Falling in \textit{Taiwei} (\textit{Taiwei} known as Supreme Palace Enclosure, contains part of Virgo, Leo, Coma Berenices, Canes Venatici and Ursa Major). (\textit{Jiutangshu.Tianwenxia} vol. 36, p. 1327)."
On 18 January, AD 774, ``a meteor as big as a \textit{Shengqi} , with tail, longer than two \textit{Zhang} from \textit{Ziwei} (\textit{Ziwei}, or \textit{Purple Forbidden Enclosure}, an ancient Chinese asterism, near Polaris), to \textit{Zhuo} (Hyades) (\textit{Xintangshu.Tianwen'er} vol.32, p.843)."
 On 9 April, AD 775, ``a meteor from the west, as big as two \textit{Shengqi}, with tail, as long as two \textit{Zhang}, to \textit{Zhuo}. (\textit{Xintangshu.tianwen'er} vol.32, p.843.) On 11 April, AD 777, a meteor as big as a peach, with a tail as long as ten \textit{Zhang}, from \textit{Hugua}  (\textit{Hugua}, or \textit{Good gourd}, an ancient Chinese asterism ) to \textit{Taiweiyuan} (\textit{Taiweiyuan}, or   \textit{Supreme Palace Enclosure} is one of the \textit{Sangong}) (\textit{Xintangshu.Tianwen'er})." In conclusion,  during AD 774-775, the only observed astronomical events are the two meteors, but they are common in other periods, and they can not be that energetic.

We extended the period, and extended the literature. We found in
\textit{ A New Catalogue of Ancient Novae and Explorations
in the History of Science} \citep[][p. 113]{Xi1955} that ``there is a fireball on the sky on January 8, AD 745", and ``at AD 827, the Arabic poet Haly and Babylonish astronomer Albumazar saw a nova as bright as a half moon lasting four months, in the tail of Scorpio". \cite{Xi1955} concluded that there was no record about novae between AD 745 to 827. \textit{Chinese Astronomical History} \citep[][p. 866]{Chen2006} recorded that there were only meteor events in AD 774-775, and the details were same as the \textit{ The General Catalogue of Chinese Ancient Astronomical Records} \citep{Bao1988}. \textit{Zizhitongjian} \citep{Sima1085} didn't show any special astronomical events had happened.
In conclusion, there was no notable astronomical event recorded in AD 774-775.

\begin{figure}
  \centering
  \includegraphics[angle=0,width=0.50\textwidth]{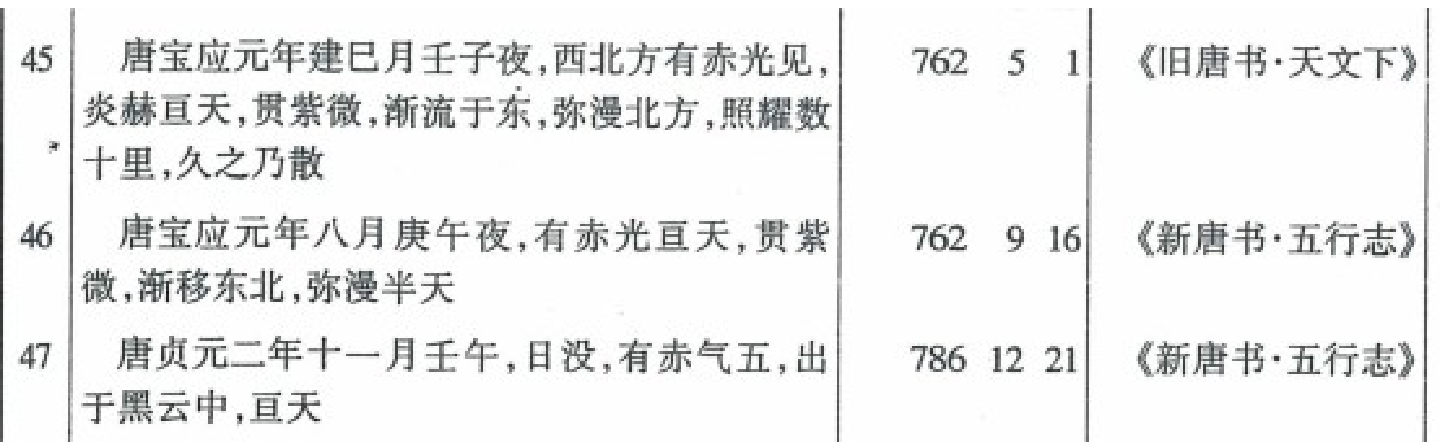}
  \caption{The records for auroras around AD 774/5 in {\it Chinese astronomical history}\citep{Chen2006} (page 951). It shows there was no remarkable  aurora event of that period, which is consistent with the {\it The General Catalogue of Chinese Ancient Astronomical Records} \citep{Bao1988}.}
  \label{fig:aurora}
\end{figure}

Then we paid special attention to the polar coronas,
as \cite{Melott2012} pointed out that a solar flare might be the source of $^{14}$C increase. If yes, we may have the records like (or even stronger than) the Carrington event \citep{Thomas2007}.
There really are records about the Carrinton event in \textit{The General Catalogue of Chinese Ancient Astronomical Records} \citep[][p. 49]{Bao1988}: on 2 September, 1859, ``red light emerged from northwest, lay across northeast, disappeared till dawn (\textit{records of Luancheng County, Hebei Province})". There were several other corona records in the same year, while for other years, the records are much rarer (as China locates relatively low latitude, only huge auroras can be seen). This indicates the Sun was active in the whole year. Consequently, there must have been much stronger polar coronae occurred
in AD 774-775. However, there is no record included in that general catalogue. The nearest two events were in AD 762 and in AD 786, as shown in figure \ref{fig:aurora}.
On 16 September, AD 762 \citep[][p. 31]{Bao1988}, ``at night,
there was red light emerge on the northwest extending the sky, reached to \textit{Ziwei}. Then moved slowly to the east, covering half of the sky (\textit{Jiutangshu.Daizongji}, vol. 11, p. 270)".
The next record
about polar corona was on 21 December, AD 786, ``at sunset, the red light came from the black clouds, covered the sky (\textit{Xintangshu.Wuxingzhi}, vol. 34, p. 894).'' There was nothing recorded
about polar corona between AD 762 and AD 786.
On the other hand, the activity of the sun often accompanied with sunspots. However, there was no Chinese ancient record about sunspots in that period reported  neither \citep{Bao1988}.
Together with the lack of sunspot records, it may indicate the solar flare was not the source.

\begin{figure}
  \centering
  \includegraphics[angle=0,width=0.5\textwidth]{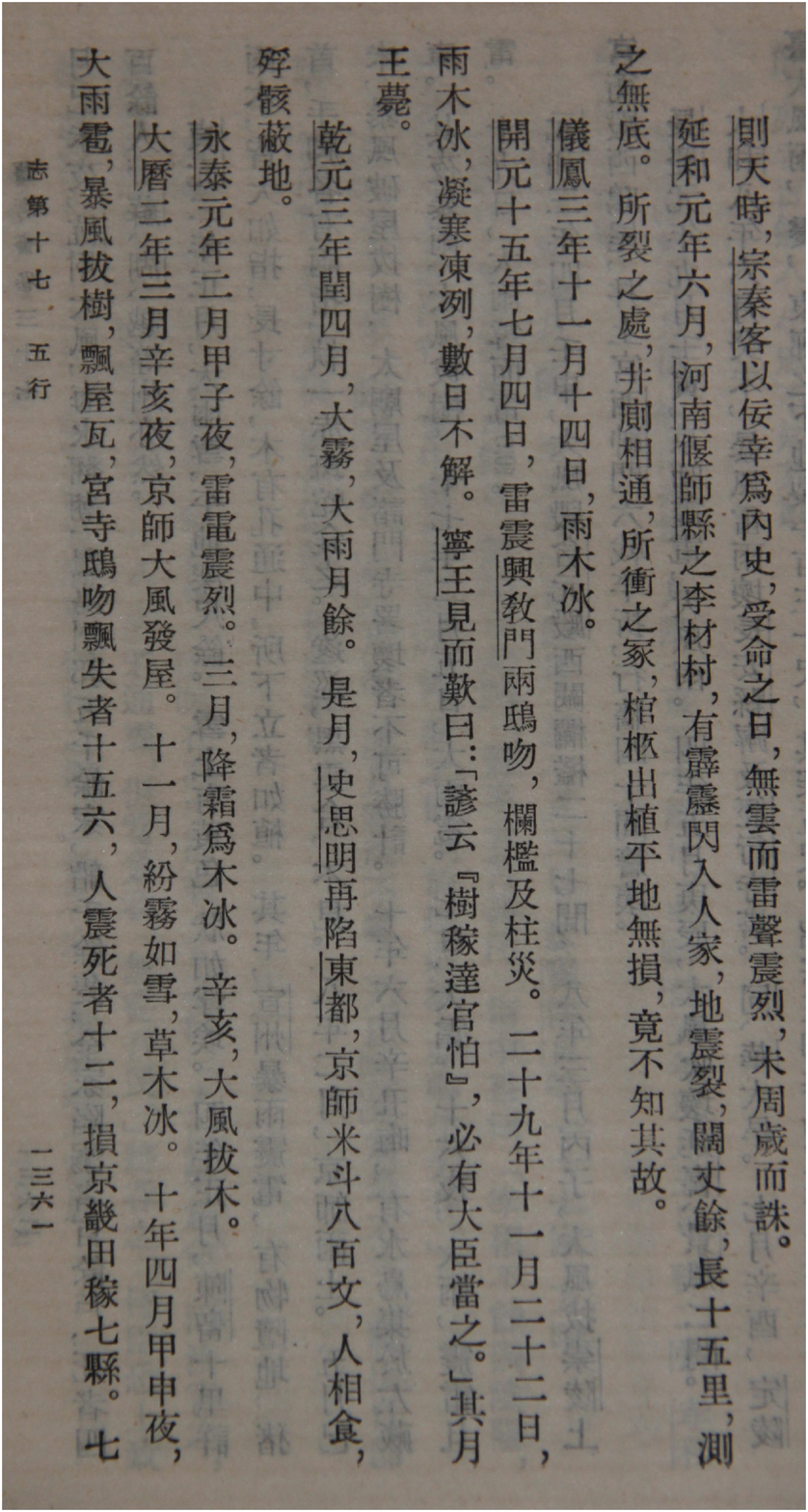}
  \caption{Records about lightening from {\it Jiutangshu} \citep{Liu945}. The left (last) two lines show the records in 774, while the other records (on the right) are much weaker than this year.}
  \label{fig:lightening}
\end{figure}

Not just restricted at the astronomical events, we also searched on different historical books describing about that period.
 Which brought our attention was that
in the \textit{Jiutangshu} (on page 1361) \citep{Liu945}: ``On 25 May, 775, there was an extremely torrential rain in the night. The strong wind uprooted the trees. The tiles of houses were blown away. 50-60\% \textit{Chiwen} (\textit{Chiwen} an ornament on the roof ridge, in the shape of a legendary animal, and was pinned to the wall by tile nails) were blown away.  20\% people died by the thunderstorm and the crops were destroyed in seven prefectures around the capital.'' This record is shown in figure \ref{fig:lightening}. There were fifteen thunderstorm and lightning events were recorded in the \textit{Xintangshu}. To show the particularity of this thunderstorm event, we list all other records during that period \citep[][on pages 941]{Ouyang1060}:

On 9 May, AD 637, ``An ancient locust tree in front of \textit{Qianyuan} temple, was shocked.''

On 9 December, AD 694, ``Thundering. The sound of the thunder is \textit{Yang}, and appeared at an inappropriate time, indicating subordinate attempting to usurp the throne.''\footnote{Notice the Chinese historical book authors did't only write down the records, but made comments occasionally.}

On 9 June, AD 704, ``Thundering. Wind uprooted the tree, and some people were killed.''

In July to August, AD 712, ``Thunderstorm and lightning struck some villagers' houses in Licai
village Yanshi prefecture, Henan province. The ground was split into a several meters wide and 7.5 km long gap, which was unfathomable. The gap made the outhouse to connect the well, or crossed the tomb, while the coffin went out of the ground without damage. The village was named Li, which is the surname of the emperor. Thunderstorm and lightning
indicate crucial punishment. Ground indicates \textit{Yin}.''

On 26 February, AD 765, ``Huge thunder in that night. There was no thunder any more until July.''

On 25 May, AD 775, ``Thunderstorm and lightning. The storm wind uprooted trees and swept tiles. Some people were killed, and the crops were destroyed in seven prefectures around the capital.'' (There is a little bit different between the \textit{Xintangshu} and the \textit{Jiutangshu})

On 20 October, AD 780, ``Thunderstorm."

 AD 783, ``\textit{Jiedushi Ge Shuzhai} (a commander) was assaulting \textit{ Jedushi Li Xilie}. When he led his army entered to \textit{Yingqiao} (name of a place), heavy rain
and thunderstorm came. 30-40\% people could not speak, and some horses and donkeys died.''

On 18 June, AD 798, ``The 1st thunder of this year.''

In AD 816, ``Thundering in the winter.''

On 28 June, AD 822, ``Strong wind, thunderstorm and lightning, dropped \textit{Chiwei}  (similar to
\textit{Chiwen}, decoration on the roof) on the imperial ancestral temple, and broke a tree.''

On 22 April, AD 834, ``Heavy rain with earthquake in \textit{Dinglingtai} (a place). The ground was splitted with twenty-six steps wide."

On 7 June, AD 843, `` Started to thunder.''

January to February in AD 864, ``Thunderstroms.''

In January of AD 876, ``Thunderstorms and hails.''

From these records, we can find the thunderstorm and lightning on 25 May, 775 was really the strongest one.

For the $^{14}$C increase event in AD 992-993 \citep{Miyake2013}, we didn't find any record about supernovae, sunspots, aurorae \citep{Bao1988, Xi1955, Chen2006} neither.

\section{Terrestrial $\gamma$-ray Flash model}

From the description above on the details of thunderstorm and lightning events in
the \textit{Xintangshu}, we can finger out that, the thunderstorm and lightning event
occurred in AD 775 led
to the heaviest damage. Though the literature just recorded the area next to the capital, considering the information could not spread easily at the ancient time, it is very possible that a much larger area was heavily destroyed by that event (or a series of thunderstorm and lightning events). We suppose that there should be an
extremely energetic thunderstorm and lightning event in AD 775, and the $\gamma$-ray counterpart might be also very energetic, which might cause the $^{14}$C increase.

Since Compton Observatory was launched in April 1991, it had recorded
numerous $\gamma$-ray flashes from the Earth \citep{Fishman1994}. The researchers who studied high-energy
atmospheric physics have found that X-rays and $\gamma$-rays can be emitted by both thunderclouds and lightings. Thunderclouds are thought to be the source of  terrestrial gamma-ray flashes (TGFs)
\citep{Joseph2012}. There are two features of TGFs: one is that the
spectra of TGFs are much harder than gamma-ray bursts, solar flares, and
other cosmic sources; the other is that their duration is short \citep{Fishman1994}. \cite{Fishman1994} estimated the energy of a typical TGF is the order of 10$^{8}$ to
10$^{9}$ ergs at a typical distance to the source of 500 km, supposing isotropic
emission.
Thunderstorms can also produce neutrons. \cite{Babich2006} estimated
that the number of neutrons yielded from a common TGF would be 10$^{15}$.
 However, the energy of a common TGFs is far less than 7$\times{}$10$^{24}$ergs$^{ }$ which can lead to the $^{14}$C content increase by
1.2\%. The influence of the neutrons yielded from a common TGF is also negligible \citep{Babich2006}. It is still possible considering the significant
increase in $^{14}$C just occurred once in 3000
years. \cite{Fuschino2011} estimated the global rate of TGFs are 220-570 TGFs per day. If we
use the upper limit 570 TGFs per day, there will be about 2$\times{}$10$^{5
}$TGFs per year. Supposing the same amount of energy of the lightning goes to the high energy $\gamma$-rays, the total energy in a usual year of the TGFs is about $10^{14}$ ergs, which is much less than the energy required.

Considering the energy of TGF obeys power law distribution, we make
a rough estimate to the rate of a huge thunderstorm and lighting event.
For a general thunderstorm and lightning event, the total energy is about $10^{9}$ erg, and the rate is around $10^{5}$ TGFs per year.
We do not have the slope of the energy distribution. For an estimate, if one needs the $^{14}$C increase comes from the TGF, the rate of an event with $7\times 10^{24}$ ergs should be $\sim 10^{-3} {\rm yr}^{-1}$. Then the slope should be $\alpha \sim -0.5$, where $\alpha$ obeys $\frac{{\rm d} N(E)}{{\rm d} E} \propto E^{-\alpha}$, while $N(E)$ is the rate of the lightning with total energy $E$. This slope is too shallow and unlikely, even though the energy ($7\times 10^{24}$ ergs) might be the total energy of all the TGFs during that year. If we consider the absorption of the atmosphere, the real energy of a typical TGF might be much larger than 10$^9$ ergs, the crisis may not be so serious.

Though we didn't find any huge destroy by the thunderstorm and lightning events at AD 992-993, this may not rule out the TGF model as the corresponding thunderstorm and lightning event can occur in the area which has no historical records.

\section{CONCLUSION and DISCUSIION}
The 1.2\% of the $^{14}$C increment within one year should have been an extremely
energetic event with $\sim 7 \times 10^{24}$ergs  $\gamma-$ray energy input on the earth. These events occurred in AD 774-775 and in AD 992-993 should have some observable counterparts, which should have been observed by the people in those decades. Chinese has long and detailed historical records in the ancient literatures. We searched for a counterpart at that period. Unfortunately, except a few meteors and comets, which are common phenomena, there are no notable astronomical records at AD 774-775 and AD 992-993. This means neither supernova, solar flare, meteorite, comet is likely to be the source.

We found the thunderstorms and lightnings were truly much intenser at AD 775, comparing with other periods. The counterpart TGF might be also intense, which may cause the $^{14}$C increase.  However, the rate of the thunderstorm (accompanied with terrestrial $\gamma$-ray flashes) with energy $\sim 7 \times 10^{24}$ergs is too low to be likely.

As the abundance of the Chinese historical records, an event with a bright optical counterpart is ruled out, which confirms \cite{Miyake2012}'s conclusion that a normal supernova is unlikely. Here we can conclude that the event which caused the  $^{14}$C increase was very likely not having an optical counterpart. Short gamma-ray burst and TGF are still candidates. According to \cite{Hambaryan2012}, a giant flare of an SGR located at 1 to 4 kpc is also possible, which may become a normal pulsar at present.

Though the records from Chinese literatures are very abundant, it is worthwhile to search the records from other countries. A similar analysis of the trees from other regions are also very helpful to confirm the global property of the $^{14}$C increase.

We would like to thank Bing Zhang, Ding-Xiong Wang, Qingwen Wu, Biping Gong, Weihua Lei, Zigao Dai and Fayin Wang for helpful discussions. This work was supported by the National Natural Science Foundation of China (Grant No. U1231101), and the National Basic Research Program (973 Program) of China (Grant No. 2014CB845800).

\label{lastpage}

\end{document}